\documentclass[onecolumn]{aa}

\usepackage[english]{babel}
\usepackage{graphicx}

\newcommand{\gsim}{\lower.7ex\hbox{$\;\stackrel{\textstyle>}{\sim}\;$}}
\newcommand{\lsim}{\lower.7ex\hbox{$\;\stackrel{\textstyle<}{\sim}\;$}}

\begin{document}

\title{Extra-planar gas in the spiral galaxy NGC\,4559}
\author{C.V. Barbieri \inst{1,} \inst{4} \and F. Fraternali \inst{2}
  \and T. Oosterloo \inst{3} \and G. Bertin \inst{4} \and
  R. Boomsma \inst{5} \and \\ R. Sancisi \inst{1} \inst{,5}}
\institute{INAF-Osservatorio Astronomico, via Ranzani 1, 40127 Bologna, Italy \and Theoretical Physics, University of Oxford, 1 Keble Road, Oxford OX1 3NP, UK \and ASTRON, PO Box 2, 7990 AA, Dwingeloo, The Netherlands
\and Universit\`a degli Studi di Milano, Dipartimento di Fisica, via Celoria 16, 20133 Milano, Italy
\and Kapteyn Astronomical Institute, University of Groningen, The Netherlands}

\date{Received / Accepted}
\abstract{We present 21-cm line observations of the spiral galaxy
NGC\,4559, made with the Westerbork Synthesis Radio Telescope. We have used them to study the H\,{\small I} distribution
and kinematics, the relative amount and distribution of luminous and dark matter in this galaxy and, in particular, the presence of extra-planar
gas.
Our data do reveal the presence of such a component, in the form of a
thick disk, with a mass of 5.9 $\times$ $10^{8}$ $M_{\sun}$ (one tenth
of the total H\,{\small I} mass) and a mean rotation velocity 25-50 km
s$^{-1}$ lower than that of the thin disk.
The extra-planar gas may be the result of galactic fountains but
accretion from the IGM cannot be ruled out.
With this study we confirm that lagging, thick H\,{\small I} layers are
likely to be common in spiral galaxies.
}

\maketitle

\section{Introduction}
A large number of spiral galaxies have been observed in the 21-cm hydrogen
line in recent years. Such observations serve two main purposes. One is to study the properties of the H\,{\small I} distribution and
kinematics, and in particular the structure of the gaseous disk, the
presence of extra-planar gas, the disk-halo connection, and the
interactions with the environment. The other is to study the
distribution of mass, the presence of dark matter, and its relationship with luminous matter.

Here we report 21-cm line observations of the spiral galaxy NGC\,4559. The main result of these observations is the discovery of an extensive system of extra-planar gas in this galaxy.

Evidence for extra-planar gas in spiral galaxies is difficult to
obtain and, to date, is available for only a few systems.  The
study of the edge-on galaxies NGC\,891 (Swaters, Sancisi, and van der
Hulst 1997) and UGC\,7321 (Matthews and Wood 2003) has revealed the
presence of neutral gas up to several kiloparsecs from the plane and a
lag in rotation with respect to the disk. A lagging H\,{\small I} layer has also been observed in NGC\,2403, a galaxy viewed at intermediate
inclination ($i = 60\degr$) (Fraternali
et al. 2002). Observations of face-on galaxies have
shown vertical motions of neutral gas frequently associated with
``holes'' in the H\,{\small I} distribution (Puche et al. 1992;
Kamphuis 1993; Boomsma et al. 2002). Vertical gradients in rotation
velocity have also been found in the ionized gas (e.g. NGC\, 5775,
Rand 2000).

These results suggest a complex gas circulation between disk and
halo in spiral galaxies: ionized gas, swept up by stellar winds
and supernova explosions, rises above the disk, cools down, and
falls back to the plane. Such a \emph{galactic fountain} (Shapiro
and Field 1976) is not the only possible interpretation for the
presence of gas outside the plane. Accretion of intergalactic
``primordial'' gas, as proposed for the High Velocity Clouds in
our Galaxy (Oort 1970), or minor mergers with small companions
(e.g. van der Hulst and Sancisi 1988) may also play an important
role.

In the present study we use the 21-cm line observations of the
spiral galaxy NGC\,4559 to investigate the structure and the
kinematics of the disk, the dark matter halo, and the presence of
the extra-planar gas. The spiral galaxy NGC\,4559 is an
isolated Scd II galaxy at a distance of 9.7 Mpc (Tully 1988) and
seen at an inclination angle of $67\degr$. It has a radially
extended H\,{\small I} layer (about 1.5 the optical size, $R_{25}$
= 15.82 kpc) and regular kinematics (see Broeils 1992).
NGC\,4559 has a fairly high star formation rate (about 1.3
$M_{\odot}$ yr$^{-1}$,  \cite{ken03}, corrected for the distance
assumed here). Various studies of NGC\,4559 have been carried out
in X-ray both with ROSAT (\cite{vog97}) and with XMM and Chandra
(\cite{cro04}). They have revealed the presence of diffuse X-ray
emission and of some Ultra Luminous X-ray sources with
luminosities L$_X \gsim $10$^{40}$ erg s$^{-1}$, probably
associated with regions of intense star formation (\cite{sor05}).

In this paper,
after a description of the H\,{\small I} observations of NGC\,4559 in
Sect.\ \ref{observation}, we derive the rotation curve (Sect.\
\ref{rotation curve}).
Then, in Sect.\ \ref{anomalous gas} we focus on the kinematically
``anomalous'' gas component and show
that such gas forms a lagging thick layer surrounding the cold
H\,{\small I} disk. In Appendix \ref{mass model} we present a mass
model for NGC\,4559, and briefly describe the properties of the dark
halo.

\section{Observations}
\label{observation}

\begin{figure}
\includegraphics[width=140mm]{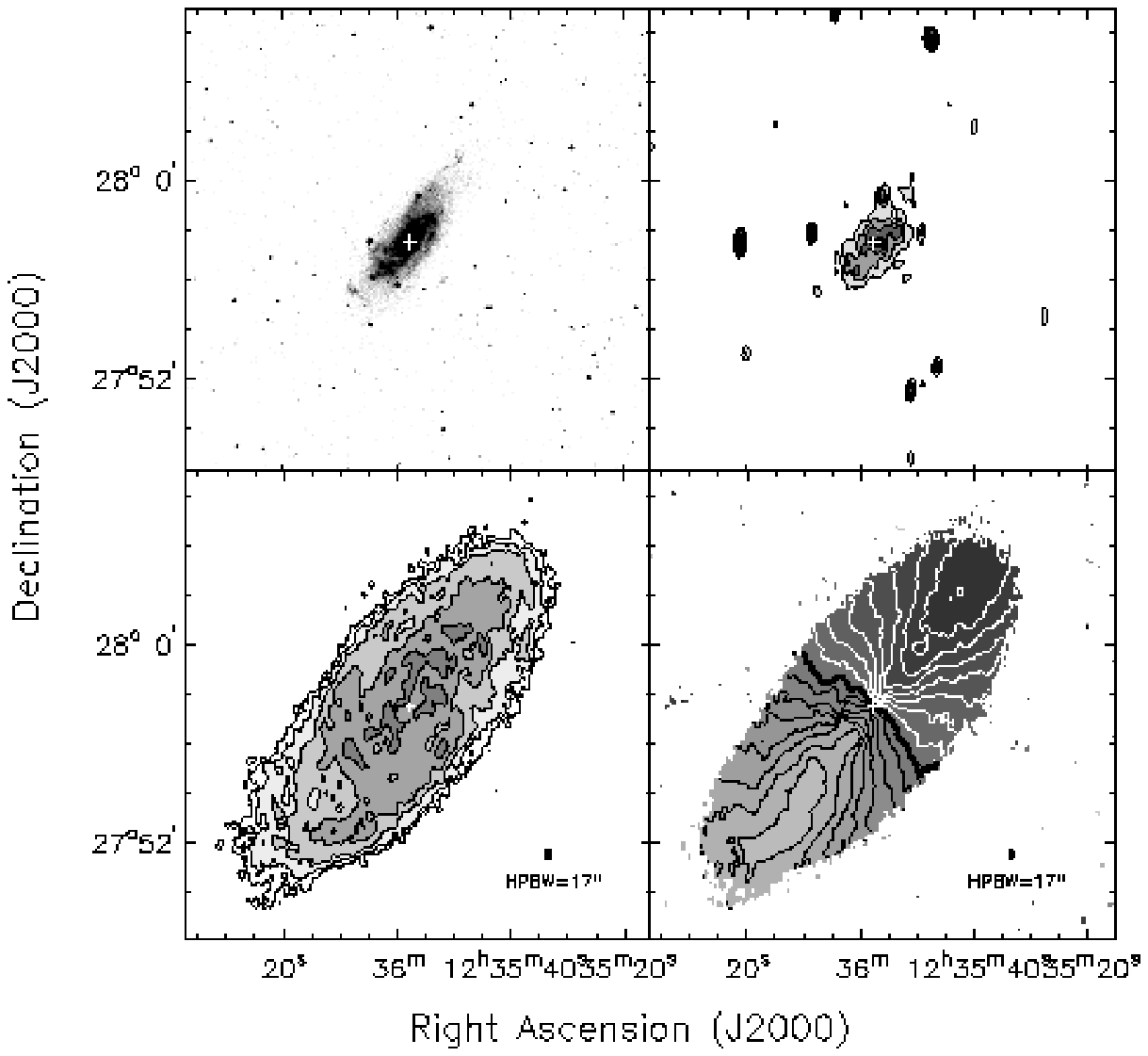}
\caption{Top left panel: optical image (DSS). Top right panel: radio continuum. The contour levels are 2, 3, 5, 6, 8, 10, 12, 14, 15 $\sigma$, with $\sigma$ = 0.15 mJy beam$^{-1}$. The column density contours in the total
H\,{\small I} image (bottom left) are 1.5, 3, 6, 12, 24 $\times$
$10^{20}$ atoms cm$^{-2}$. Bottom right panel: velocity field at
17$\arcsec$ resolution. In the velocity field, contours are
separated by 15 km s$^{-1}$; the thick line marks the systemic velocity
(810 km s$^{-1}$). The receding side is darker. The scale factor is 1$\arcmin$
= 2.8 kpc. A small cross indicates the
kinematical centre of the galaxy.}
\label{pres}
\end{figure}

The DSS (Digitized Sky Survey) image of NGC\,4559 is
presented in Fig. \ref{pres}. The optical disk shows a
multiple-armed spiral structure. The South-West side is the near
side, assuming that the spiral arms are trailing. The main optical
and radio parameters are summarized in Table \ref{ngc4559}. We
have observed NGC\,4559 with the Westerbork Synthesis Radio
Telescope (WSRT). The WSRT observing parameters are summarized in
Table \ref{observ}.

\subsection{Data reduction}
Calibration and
data reduction have been performed using the standard procedures
of the MIRIAD (Multichannel Image Reconstruction, Image Analysis
and Display) package. The MIRIAD task UVLIN has been
used to derive the radio continuum emission and to subtract it from
the line channels. This was done by interpolating with a straight line between the channels free of line emission at both ends of the band. Using all baselines we
have obtained the full resolution (17$\arcsec$ = 790 pc) data
cube. We have applied a Hanning smoothing to these data. The velocity resolution is 8.2 km $s^{-1}$. We have also produced data cubes at resolutions of 26$\arcsec$ and 60$\arcsec$.
The raw images have been deconvolved with the CLEAN algorithm (Clark
1980). A summary of the parameters for the three
cubes is given in Table \ref{cube}.

\begin{flushleft}
\begin{table}[ht]
\centering
\begin{tabular}{lcr}
\hline
\hline
Parameter & NGC\,4559 & Ref.\\
\hline
\hline
Morphological type & Scd &4\\
Luminosity class & II&6\\
Optical centre \\($\alpha$,$\delta$ J2000) & 12$^h$35$^m$57$^s$.6$~~$+27$\degr$57$\arcmin$31$\arcsec$.4& 2\\
Kinematical centre\\ ($\alpha$,$ \delta$ J2000) & 12$^h$35$^m$58$^s$$\pm$7$^s$~+27$\degr$57$\arcmin$32$\arcsec$$\pm$5$\arcsec$&7\\
Distance (Mpc) &9.7 ~~(1$\arcmin$=2.8 kpc)&1\\
$L_{\rm B}$(L$_{\sun}$) & 1.06 $\times$10$^{10}$&4 \\
$L_{\rm K}$(L$_{\sun}$) & 2.53 $\times$10$^{10}$&4 \\
Disk scale length (kpc) &1.9&5\\
$R_{25}$(kpc) &15.82 &1\\
$R_{\rm Ho}$(kpc) &16.24&3 \\
Systemic velocity (km s$^{-1}$)& 810 $\pm$ 4&7\\
Mean H\,{\small I} inclination angle (deg)&  67.2 $\pm$ 0.6&7\\
Mean position angle (deg)& -37.0 $\pm$ 1.4 & 7\\
\hline
\end{tabular}
\caption{Optical and radio parameters for NGC\,4559. (1) Tully (1988);
  (2) Karachentsev and Kopylov (1990); (3) Holmberg (1958); (4) Gavazzi
  and Boselli (1996); (5) 2MASS catalog; (6) RC2 catalog; (7) This work.}
\label{ngc4559}
\end{table}
\end{flushleft}

\begin{table}[ht]
\begin{center}
\begin{tabular}{lc}
\hline
\hline
                                & {NGC~4559}\\
\hline
\hline
Observation date               & 2001 Feb 01; June 29 \\
Length of observation               &2 $\times$ 12 h\\
Number of antennas                                   & 14 \\
Baseline (min-max-incr) & 36-m 2700-m 36-m\\
Pointing\\
 ($\alpha$,$\delta$ J2000) &  $12^h35^m57^s.7$$~$ +27$\degr$57$\arcmin$35$\arcsec$.99\\
Central velocity (km s$^{-1}$)  &594.4 \\
Central frequency (MHz)                          &  1419.747\\
Total bandwidth (MHz)                               & 2458 \\
Total bandwidth (km s$^{-1}$)                         & 525 \\
Number of channels                                     & 128 \\
Channel separation (kHz)                        & 19.2 \\
Channel separation (km s$^{-1}$)                 & 4.1\\
\hline
\end{tabular}
\end{center}
\caption{Observation parameters.} \label{observ}
\end{table}

\begin{table}[ht]
\begin{center}
\begin{tabular}{lccr}
\hline
\hline
                                && {NGC~4559}\\
\hline
\hline
 &17$\arcsec$ & 26$\arcsec$ &60$\arcsec$\\
\hline
\hline
HPBW ($\arcsec$) & {12.2 x 24.5} & {22.7 x 30.9} &  {60.0 x 60.0} \\
HPBW (kpc)& {0.57 x 1.14} & {1.05 x 1.44}& {2.8 x 2.8}\\
P.A. of synthesized beam (deg)& 0.2 & 1.0& 1.0\\
R.m.s. noise per channel (mJy beam$^{-1}$) & 0.52 & 0.48 & 0.68 \\
R.m.s. noise per channel (K) & 1 & 0.4 & 0.1\\
Minimum detectable column density (5 $\sigma$):\\
per resolution element (atoms cm$^{-2}$)&7.73 $\times 10^{19}$&3.0 $\times 10^{19}$&0.83 $\times 10^{19}$\\
per resolution element ($M_{\sun}$ pc$^{-2}$)&0.63&0.24&0.06\\
Conversion factor (K mJy$^{-1}$) & 2 & 0.86 & 0.17 \\
\hline
\end{tabular}
\end{center}
\caption{Parameters of the data cubes.} \label{cube}
\end{table}

\subsection{H\,{\small I} distribution and kinematics}
\label{data description}
We have analyzed the data cubes using the
Groningen Image Processing System (GIPSY).

\begin{figure}
\includegraphics[width=120mm]{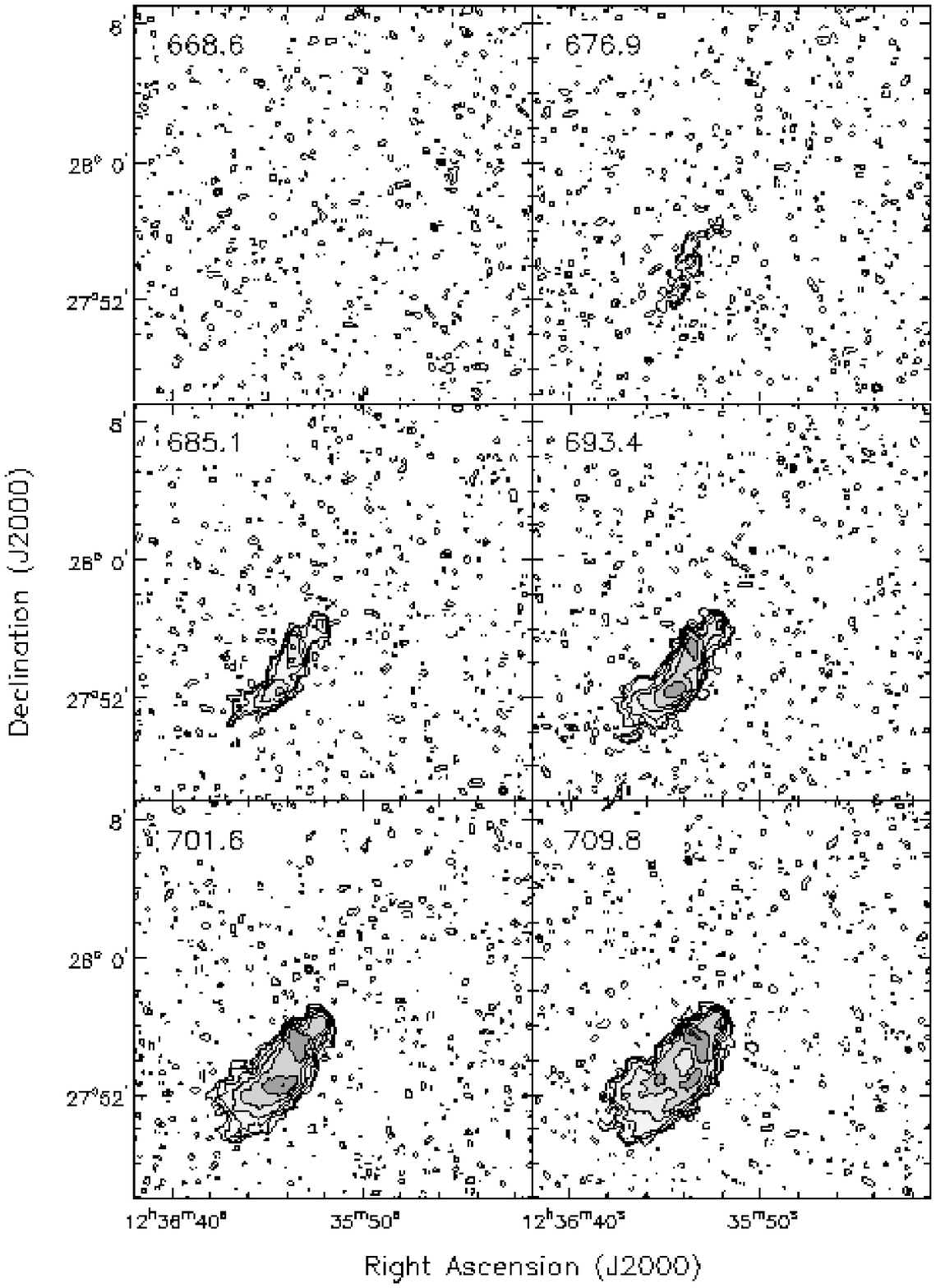}
\caption{H\,{\small I} channel map at 26$\arcsec$ resolution. The
heliocentric radial velocity (km s$^{-1}$) is shown in the upper left
corner. Contours are -2, 2, 4, 8, 16, 32, 64 $\sigma$, with $\sigma$ = 0.48 mJy beam$^{-1}$. A small cross indicates the
kinematical centre of the galaxy.}
\label{chan30}
\end{figure}

We have obtained total H\,{\small I} images at all three resolutions by
adding the channel maps containing line emission (from 688.6 km
s$^{-1}$ to 957.2 km s$^{-1}$). In each channel map, we
have defined the region of the H\,{\small I} emission by applying  masks to the
data. The masks have been obtained by smoothing the 26$\arcsec$ resolution
data cube and applying a cutoff of 2 $\sigma$. Figure \ref{pres} shows
the total H\,{\small I} image and the velocity field at
17$\arcsec$ resolution. This has been obtained by fitting, at each
position, a Gaussian to the line profiles. Figure \ref{chan30} gives the H\,{\small I} channel maps at 26$\arcsec$ resolution.

\begin{figure}
\includegraphics[width=120mm]{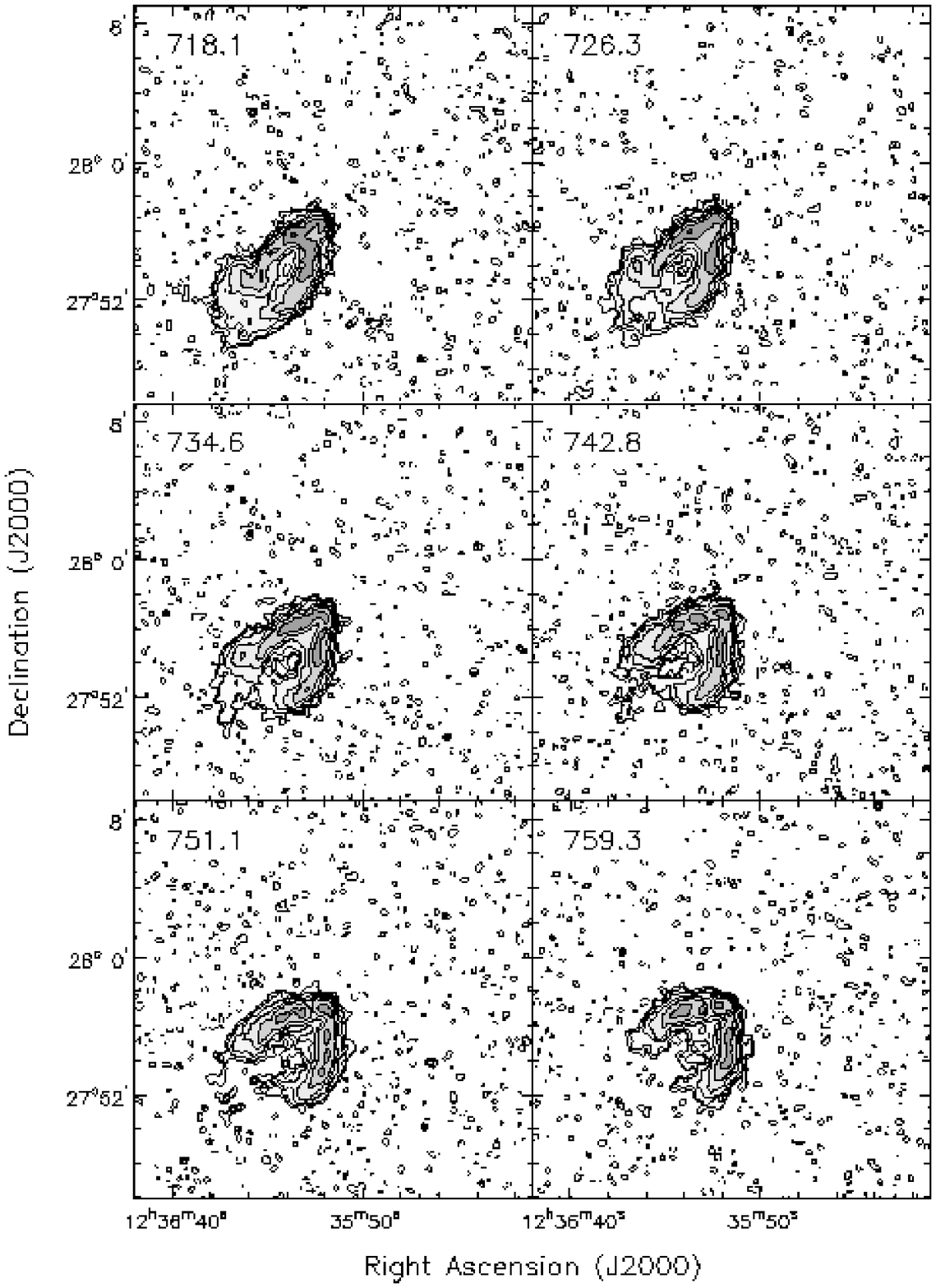}
\centering{FIG.\ref{chan30}-\emph{Continued}}
\end{figure}

\begin{figure}
\centering
\includegraphics[width=120mm]{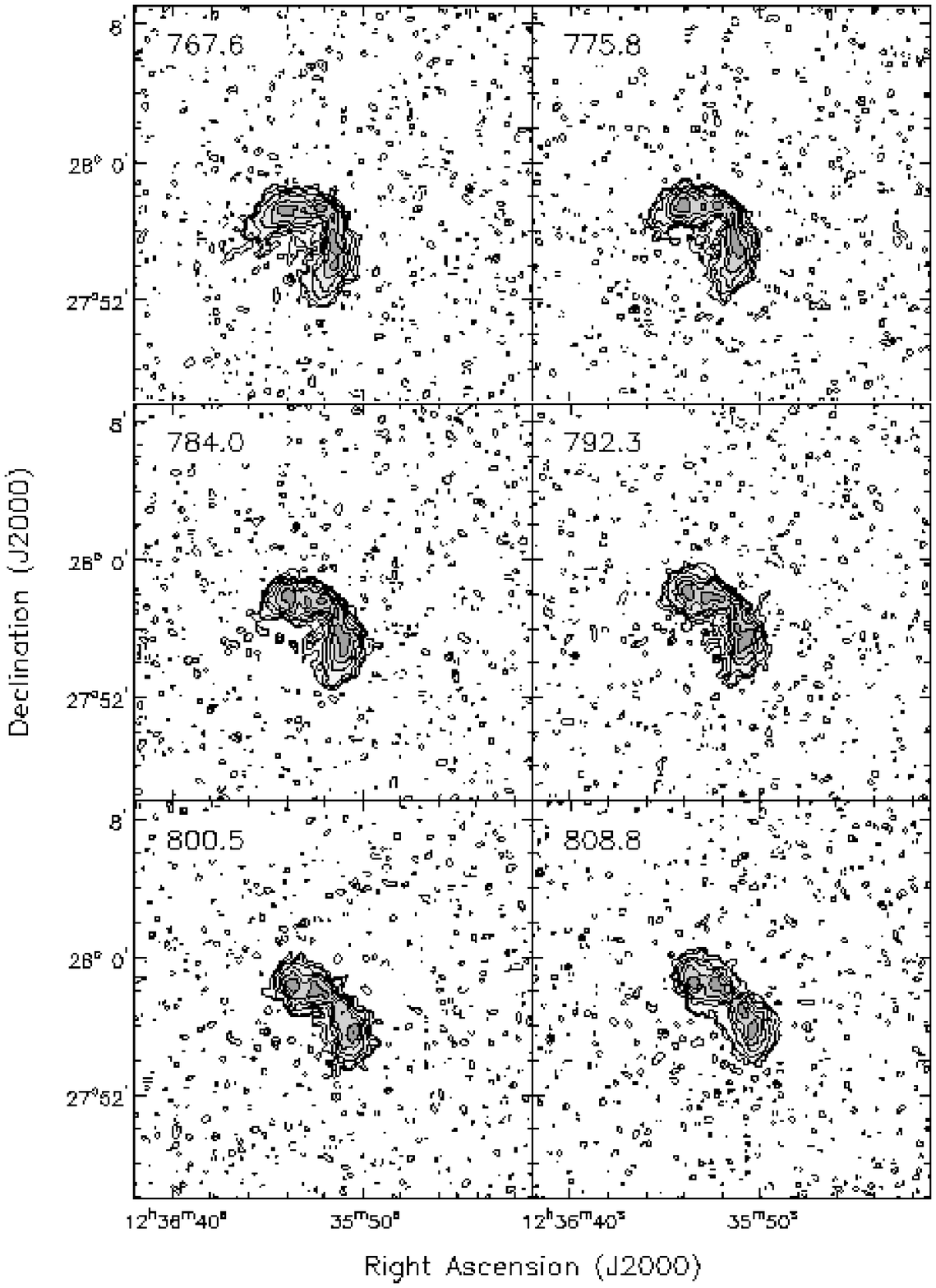}
\centering{FIG.\ref{chan30}-\emph{Continued}}
\end{figure}

\begin{figure}
\includegraphics[width=120mm]{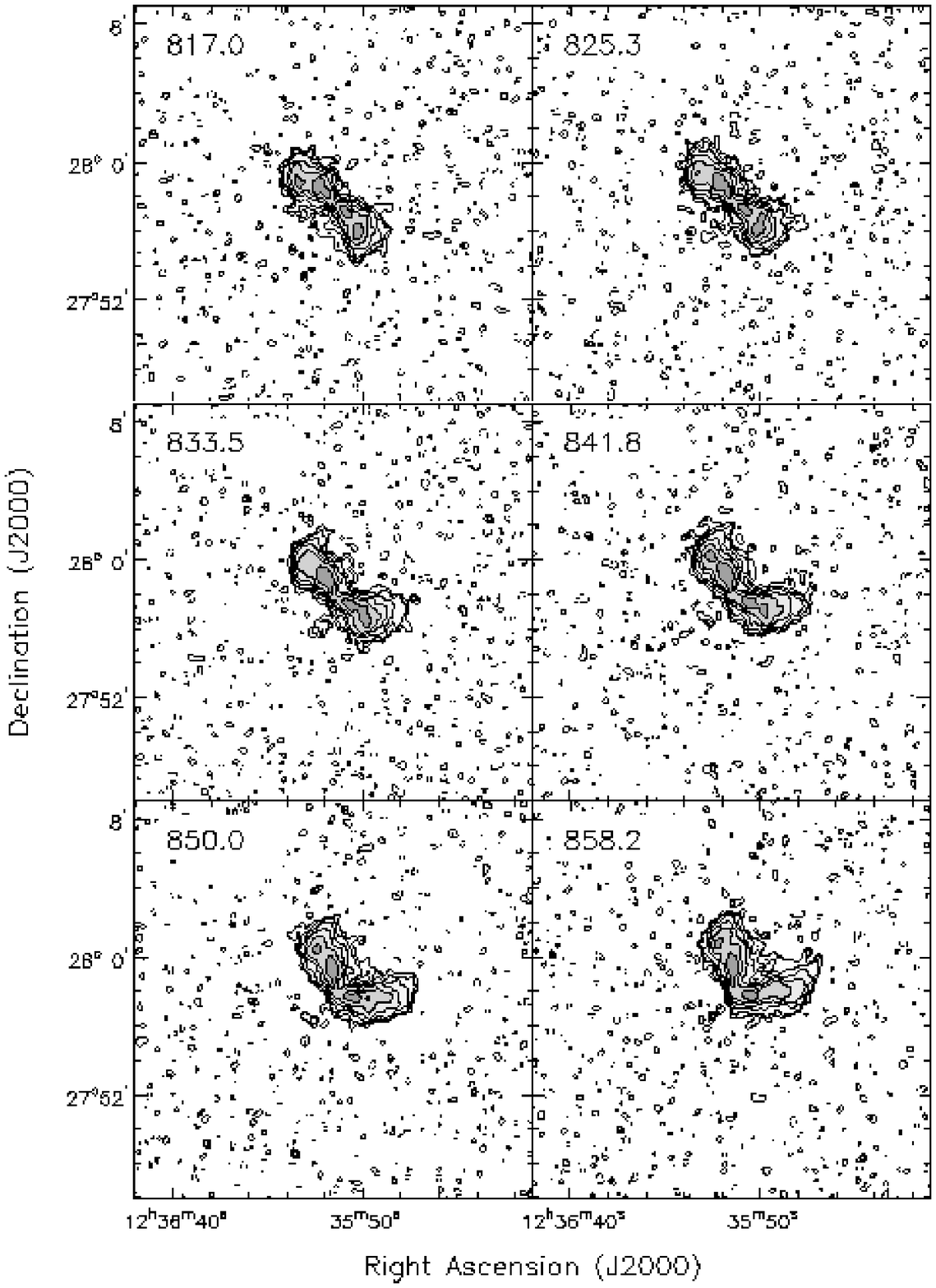}
\centering{FIG.\ref{chan30}-\emph{Continued}}
\end{figure}

\begin{figure}
\includegraphics[width=120mm]{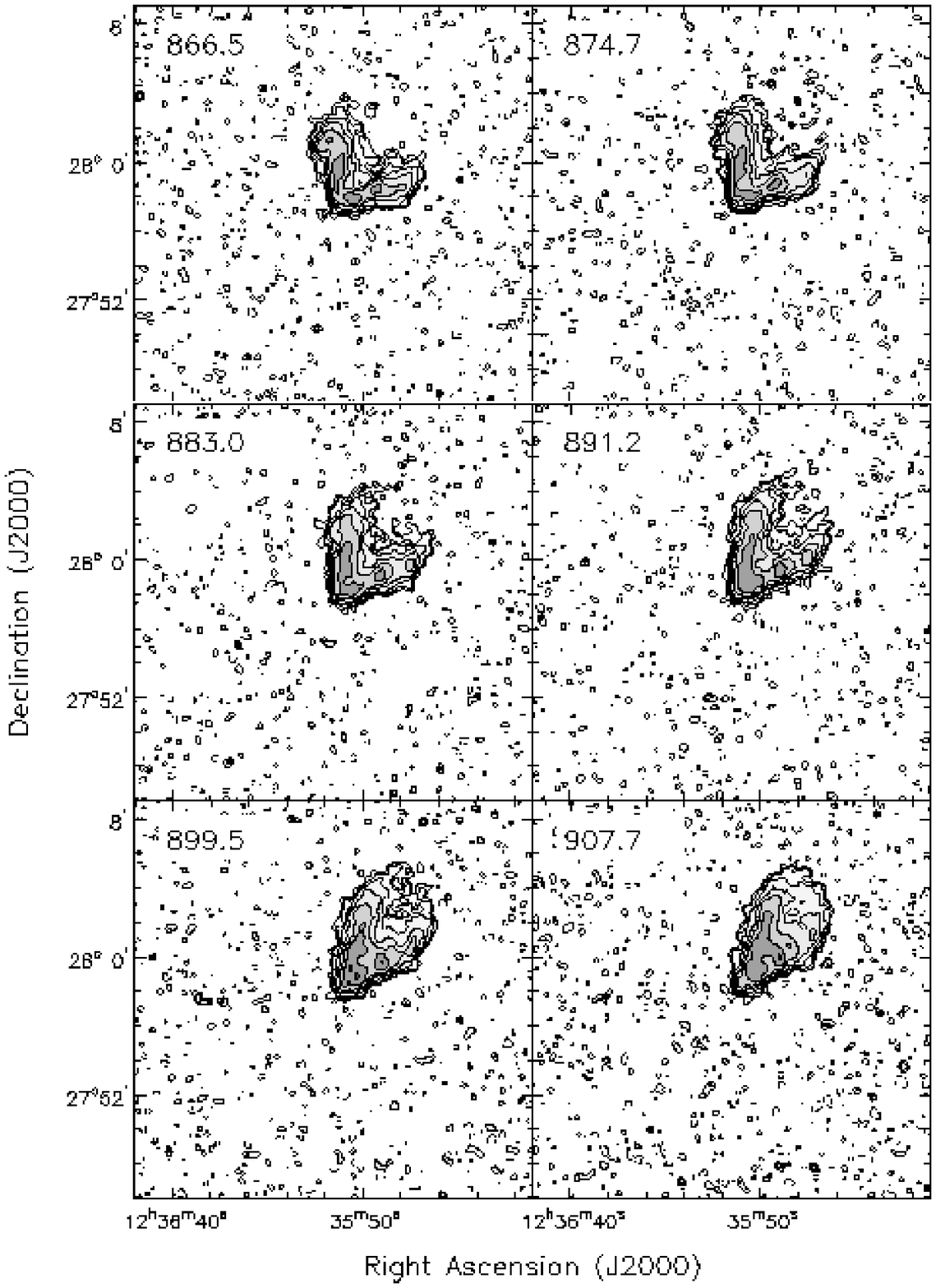}
\centering{FIG.\ref{chan30}-\emph{Continued}}
\end{figure}

\begin{figure}
\includegraphics[width=120mm]{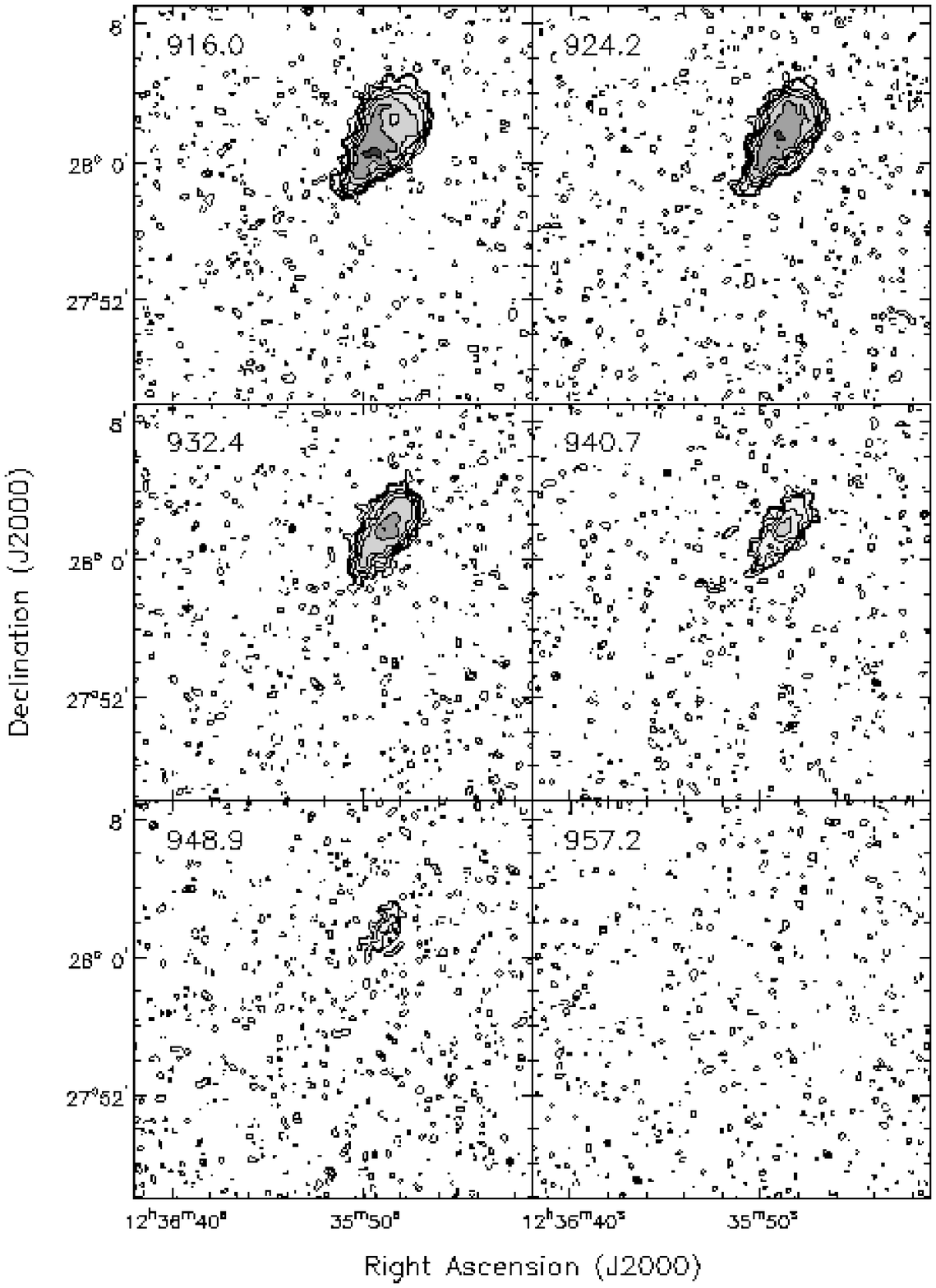}
\centering{FIG.\ref{chan30}-\emph{Continued}}
\end{figure}

\begin{figure}
\includegraphics[width=140mm]{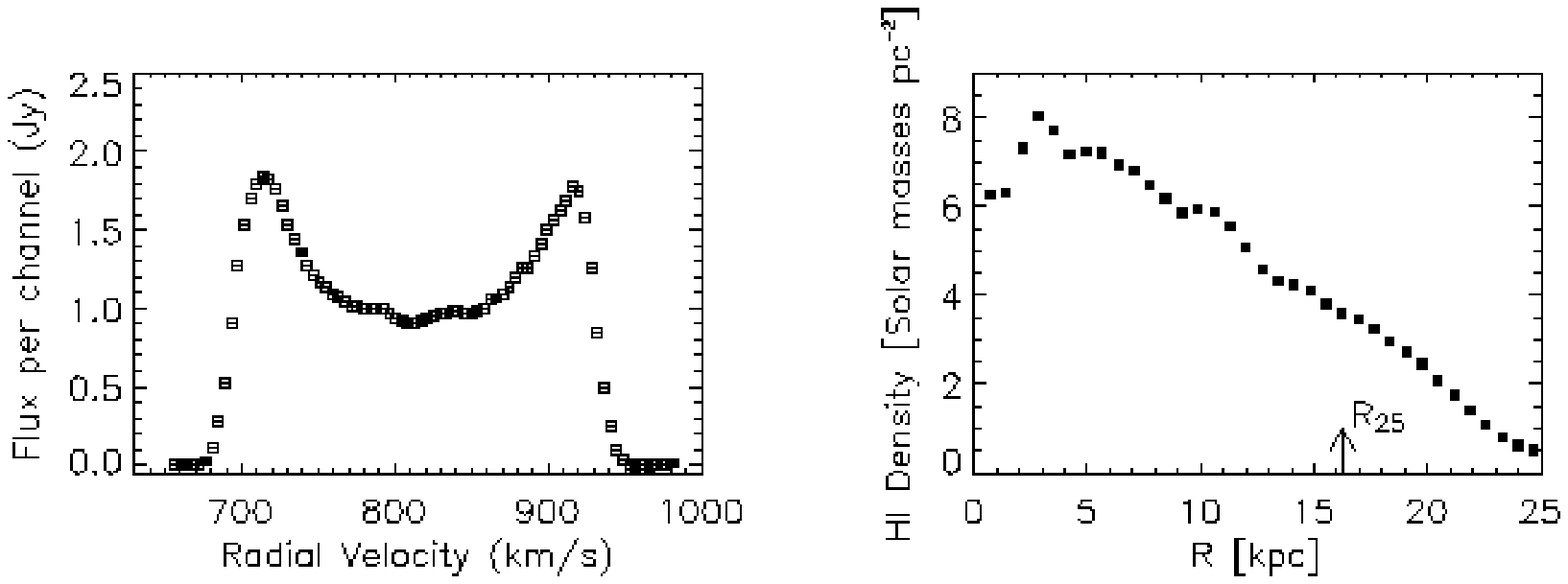}
\caption{Global H\,{\small I} line profile (left) and column
density radial profile (right).} \label{global_rev}
\end{figure}

Figure \ref{global_rev} (left panel) shows the global
H\,{\small I} line profile obtained by adding the flux densities
for each channel map. The H\,{\small I} mass of NGC\,4559,
corrected for the primary beam attenuation, is (6.75 $\pm$ 0.02)
$\times 10^9 M_{\sun}$. This agrees with the value of 6.72 $\times
10^9 M_{\sun}$ obtained by Broeils (1992) and of 6.4 $\times 10^9
M_{\sun}$ derived by Shostak (1975) with the NRAO (National Radio
Astronomical Observatory) 300-foot (91m) telescope. The right 
panel of Figure \ref{global_rev} shows the radial profile of the
H\,{\small I} column density for NGC\,4559 obtained by averaging
the column densities in ellipses with position
 and inclination angles as reported in Table 1.
Note the depression in the centre  (R$<$2.5 kpc) and the linear
decline in the outer regions.

The H\,{\small I} density distribution and kinematics of NGC\,4559
are not symmetric on the two sides of the galaxy. The lopsidedness
in the density distribution can be seen in Figure 1 (bottom left
panel). The disk is more extended on the approaching side (S-E),
reaching out to 26.6 kpc from the centre, while on the receding
side it reaches 22.4 kpc. These are the distances, along the major
axis, from the centre of the galaxy to the last column density
contour of 0.6 $M_{\sun}$ pc$^{-2}$. The S-E part is warped (see
channels from 685.1 km s$^{-1}$ to 734.6 km s$^{-1}$). The warp is
also visible in the total H\,{\small I} image and in the velocity
field (Fig. \ref{pres}). In Sect.~\ref{rotation curve} we derive
separate rotation curves for the two halves of the galaxy and show
that they differ significantly, i.e.\ the galaxy is lopsided also
in its kinematics . This can be seen in the velocity field as a
different shape of the iso-velocity contours in the northern and
the southern sides. Despite these asymmetries in the H\,{\small I}
distribution and kinematics, the global H\,{\small I} line profile
shown in Figure \ref{global_rev} is highly symmetric. This is a
surprising result of the combined effects of lopsidedness in
kinematics and in density distribution. In the past, studies of
the occurrence of lopsidedness in spiral galaxies have been based
on the shape of the global line profiles (\cite{ric94, hay98}).
The present result indicates that such studies are likely to have
missed a significant fraction of galaxies that, like NGC\,4559,
are lopsided, despite the symmetry of their global line profiles.

The position-velocity (p-v) diagrams parallel to the major and
to the minor axes (Fig. \ref{mod_cod} and Fig. \ref{models}) show that the H\,{\small I} line
profiles are asymmetric with respect to the peak. In
Fig. \ref{mod_cod} (left) the rotation curve (white dots) follows the ridge of the H\,{\small I} emission and broad, low-level extensions are visible toward the systemic velocity. At positions close to the galaxy centre (0$\arcmin$ to
$+$2$\arcmin$.5), there are traces of emission at ``forbidden'' velocities
($2^{nd}$ quadrant) differing by about 150 km s$^{-1}$ from
the rotation velocity.
This ``forbidden H\,{\small I}'' is also visible in the individual channel maps
between 817 km s$^{-1}$ and 874.7 km s$^{-1}$, southeast of the
centre. The asymmetric line profiles appear similar to those observed
in NGC\,2403 (Fraternali et al. 2002). They suggest the existence of
an H\,{\small I} component other than the thin, regularly rotating
disk (cf. Fig. \ref{mod_cod} left and right panels).

The H\,{\small I} distribution shows several holes. The most
remarkable is located at $\alpha$ = 12$^h$36$^m$4$^s$ $\delta$ = 27$\degr$57$\arcmin$7$\arcsec$ (see Sect. \ref{holes}).

\section{Rotation curve}
\label{rotation curve}

We have used the velocity field shown in Fig. \ref{pres}
to obtain the kinematical parameters and the rotation curve of NGC\,4559. For this we have followed the standard tilted-ring fitting procedure described by Begeman (1987). The rings have been chosen with a radial
increment corresponding to the beamwidth of the observations (17$\arcsec$).
Points have been weighted by the cosine of the azimuthal angle
with respect to the major axis. To determine the position of the
kinematical centre and the systemic velocity, only rings within
22.4 kpc from the galaxy centre have been considered. The mean values
of the centre position and of the systemic velocity are reported
in Table \ref{ngc4559}.

We have then proceeded to determine the inclination angle. In this
procedure, the two sides of the galaxy are analyzed separately.
The H\,{\small I} distributions on the two sides of the galaxy
(see Sect. \ref{data description}) are different. We have fitted
the tilted rings out to 22.4 kpc on the receding side and out to
26.6 kpc on the approaching side. We have obtained mean
values for the inclination angle in the approaching and receding
sides by averaging such values between R=5 kpc and R=17 kpc.
Outside this range the inclination angle is not constant. We found
i$=$66.1$\degr$ $\pm$ 0.8$\degr$ and i$=$68.4$\degr$ $\pm$
0.6$\degr$ for the approaching and receding sides respectively.
Finally, we have determined the position angles separately for the
two sides of the galaxy. The mean values (obtained by averaging
over the whole galaxy) for the inclination and the position angles
are 67.2$\degr$ $\pm$ 0.6$\degr$ and -37.0$\degr$ $\pm$
1.4$\degr$, respectively. 

\begin{figure}
\includegraphics[width=100mm]{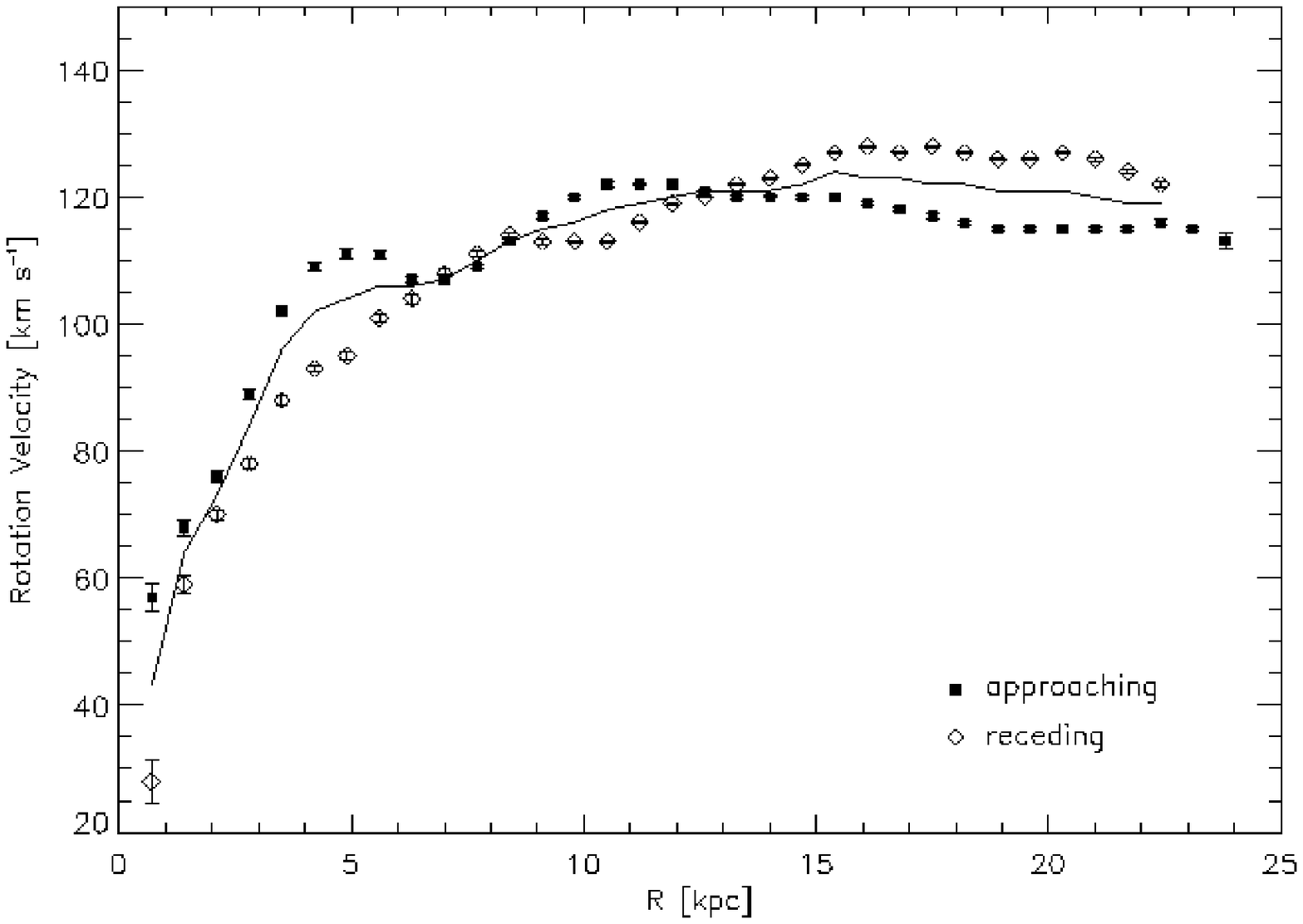}
\caption{Rotation curves obtained for the two sides of NGC\,4559.
The continuous line shows the curve obtained for the whole galaxy.}
\label{rotcur}
\end{figure}

Using the values found for the inclination and position angles, we
have determined the rotation curve first for the two sides of the
galaxy separately and then for the whole galaxy (Fig.
\ref{rotcur}). As already noted, NGC\,4559 is kinematically
lopsided. The rotation curves of the two sides of the galaxy are
significantly different. The rotation curve of the approaching
side rises to 110 km s$^{-1}$ inside 5 kpc and subsequently
flattens, while on the receding side it continues to rise out to
the outer parts where it reaches 123 km s$^{-1}$. This can also be
seen in the velocity field, where the isovelocity contours are
more curved on the approaching S-E side (see Figure 1).

In the Appendix we present two mass models for NGC\,4559. The
maximum disk fit gives a mass-to-light ratio M/L=0.27 in K-band.
In order to determine the error bars of the rotation curve in
Fig.\ref{rotcur_anom} and Fig. \ref{isot}, we have
 constructed a residual velocity field by subtracting a model from
the observed velocity field.
Then, we have taken the mean values along ellipses defined by the position
and the inclination angles obtained from the tilted-ring model analysis.
In this way the error bars give also a measure of the deviations from
circular motion.

\section{Extra-planar gas}
\label{anomalous gas}

\begin{figure}
\centering
\includegraphics[width=140mm]{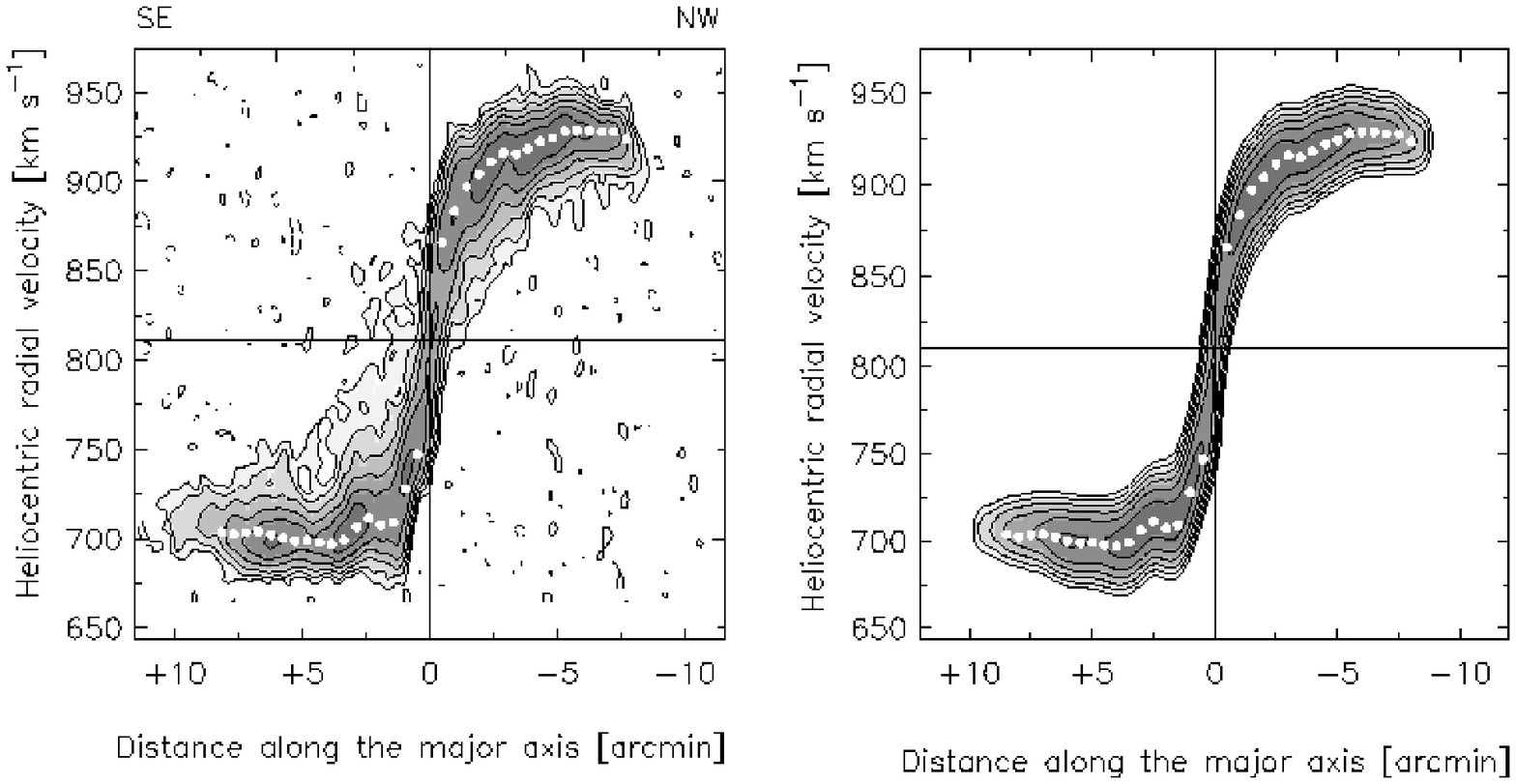}
\caption{Left: H\,{\small I} p-v diagram at 26 $\arcsec$
resolution along the major axis of NGC\,4559 (P.A. = -37$\degr$, V$_{sys}$ =
810 km s$^{-1}$). Right: model p-v diagram along the
major axis for a thin H\,{\small I} disk. Contours are -2, 2, 4, 8,
16, 32, 64 $\sigma$, with $\sigma$ = 0.48 mJy beam$^{-1}$. The white dots show
the (projected) rotation curve.} \label{mod_cod}
\end{figure}

The H\,{\small I} position-velocity diagram along the major axis
(Fig. \ref{mod_cod}, left panel) shows the presence of systematic asymmetries in the form of wings in the line profiles away from
the rotation velocity (white dots) and towards the systemic
velocity. The model of a thin H\,{\small I} disk with a Gaussian velocity profile (cold disk), as shown in the right hand panel of
the same figure, does not have such wings. We call the gas responsible
for these wings ``anomalous'' gas. The presence of anomalous gas is
more striking in the approaching than in the receding part of the
galaxy; such a difference is probably related to the kinematic
lopsidedness of the cold disk (see Sect. \ref{origin}). A similar anomalous gas component was found in NGC\,2403 (Fraternali et al. 2002) and was interpreted as
an extra-planar component that forms a thick H\,{\small I}
layer embedding the thin disk.
Most likely the two components do not coexist in the region close to the equatorial plane.

\subsection{Distribution and kinematics of the anomalous H\,{\small I}}
In order to study  the distribution and kinematics of the anomalous
H\,{\small I}, we have separated it from the regular H\,{\small I} component.
We have assumed that the thin cold disk contribution to the line profile is described by a Gaussian function centered on
the rotation velocity of the galaxy and we have defined as
``anomalous gas'' the H\,{\small I} emission not included in the Gaussian profile.
The cold disk has thus been modeled by
fitting a Gaussian function to the line profile, after applying a clip
to the data at the level of 10\% of the peak value. The obtained
velocity dispersions are in the range 8-12 km s$^{-1}$. The resulting
Gaussian model cube has been subtracted from the data, generating a
new cube made only of the residual anomalous gas.
This has all been done with the 17$\arcsec$ data cube.
Subsequentely, the residual cube has been smoothed to 60$\arcsec$
resolution to improve the S/N ratio.

\begin{figure}
\includegraphics[width=140mm]{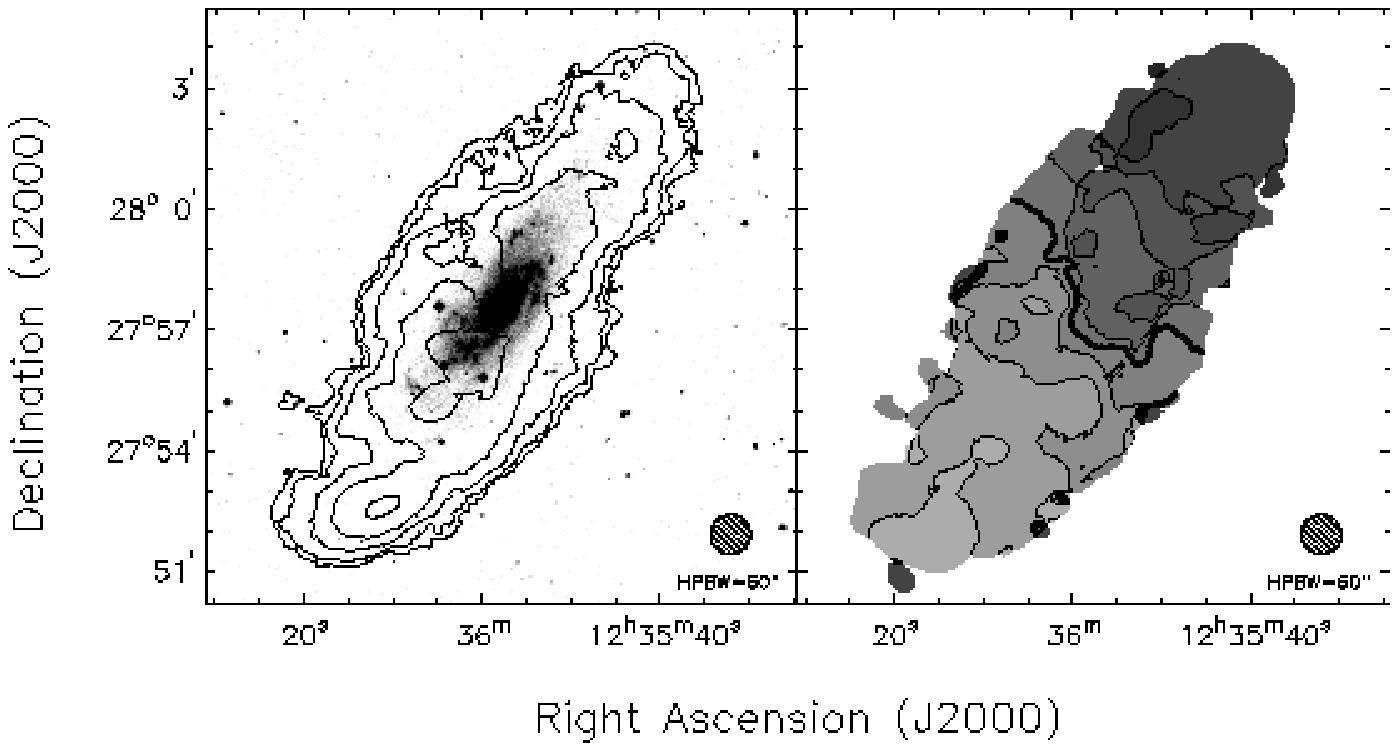}
\caption{Total H\,{\small I} image (left) and intensity-weighted mean velocity field (right)
for the anomalous gas in NGC\,4559. The column density contours are
1, 2, 4, 8, 16, 32 $\times$ $10^{19}$ atoms cm$^{-2}$. The left panel shows the
density distribution of the anomalous gas overlaid on the optical
DSS image of the galaxy. The right panel shows the velocity field,
with isovelocity contours separated by 25 km s$^{-1}$; the thick line
corresponds to
 the systemic velocity.} \label{anom}
\end{figure}

\begin{figure}
\includegraphics[width=100mm]{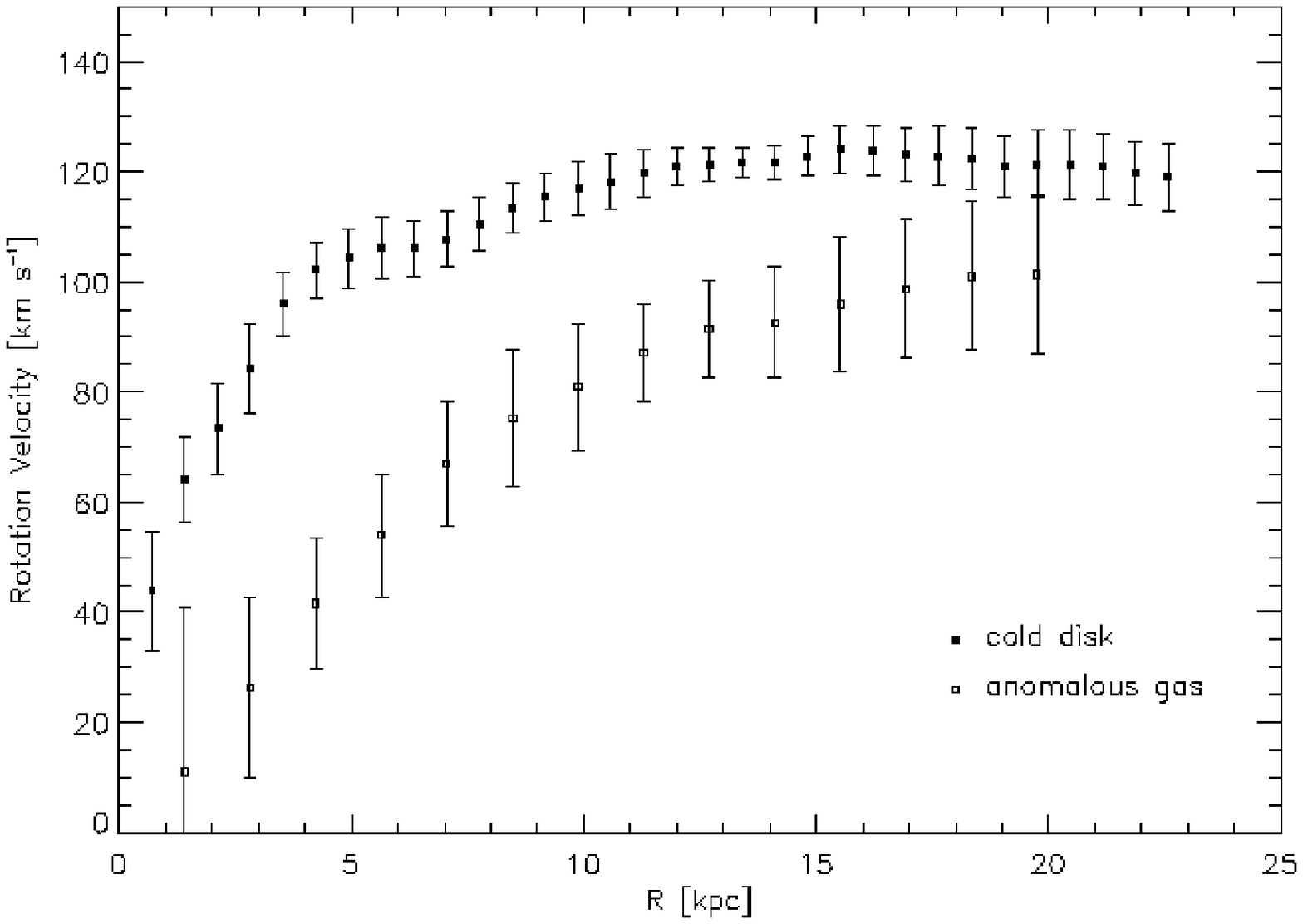}
\caption{ Rotation curves for the cold disk (filled symbols, same as in Figs. \ref{rotcur} and \ref{isot}, 17$\arcsec$ data), and for the anomalous gas (open symbols). The curve for the anomalous gas has been obtained from the 60$\arcsec$ data.}
\label{rotcur_anom}
\end{figure}

The left panel of Fig. \ref{anom} shows the distribution of the
anomalous gas overlaid on the optical image.
Its diameter is 43 kpc and the total mass is about 5.9 $\times$
$10^{8}$ $M_{\sun}$ (that is, about 1/10 of the total H\,{\small I}
mass).
The velocity dispersion of this anomalous gas, obtained from the 3D
models discussed below (see Sect. \ref{3-D models}), is about 12-25 km
s$^{-1}$.
The right panel of the same figure presents the intensity-weighted
mean velocity field of the anomalous gas.
Much like for the cold disk, the kinematics of the anomalous gas
is dominated by differential rotation.

The spatial distribution of the anomalous gas is quite
homogenous and isotropic (but see Section 5.2). This is only
partially due to the low resolution of the image in Figure 6 and
suggests that the anomalous gas is not the result of a single
recent local event (such as the capture of a lump of gas) but
instead of a widely spread phenomenon. Moreover, the kinematics of
the anomalous gas is regular and closely follows that of the thin
disk. This suggests a close relation with the thin disk, as in a
galactic fountain (see Section 5.3).

We have fitted the tilted ring model to the velocity field of the
anomalous gas in order to obtain the rotation curve. This has been
done by setting all the relevant  geometrical parameters fixed to
the values found for the cold disk. The uncertainties in the
derivation of the rotational velocities (error bars in Fig.
\ref{rotcur_anom}) have been estimated from the residual velocity
field, by taking the mean values along ellipses as done for the
cold disk (see Sect. \ref{rotation curve}). Figure
\ref{rotcur_anom} shows the rotation curves of the cold disk and
of the anomalous gas. The difference is of about 20 km s$^{-1}$ in
the outer regions and 60 km s$^{-1}$ in the inner parts. Table
\ref{cold-anom} summarizes the properties of the cold disk and of
the anomalous gas.

\begin{table*}[ht]
\begin{tabular}{lcr}
\hline
\hline
 Parameter & Cold Disk&Anomalous Gas\\
\hline
\hline
Mass ($M_{\sun}$)& 6.2 $\times$ $10^{9}$&5.9 $\times$ $10^{8}$\\
Total extent (kpc)&49&43\\
Scale height (kpc)&0.2&$\leq$4~$^{\rm 1}$\\
Maximum rotation velocity (km s$^{-1}$)&123&102\\
Mean velocity dispersion (km s$^{-1}$)&8-12~$^{\rm 2}$&12-25~$^{\rm 2}$\\
\hline
\end{tabular}
\caption{Properties of the cold disk and of the anomalous
gas. $^{1,2}$ Values obtained from the 3D model. $^{2}$ The mean
velocity dispersion ranges from the lower value in the outer parts to
the upper value in the inner parts of the galaxy.}
\label{cold-anom}
\end{table*}

\subsection{Models}
\label{3-D models} 
\begin{figure}
\includegraphics[width=140mm]{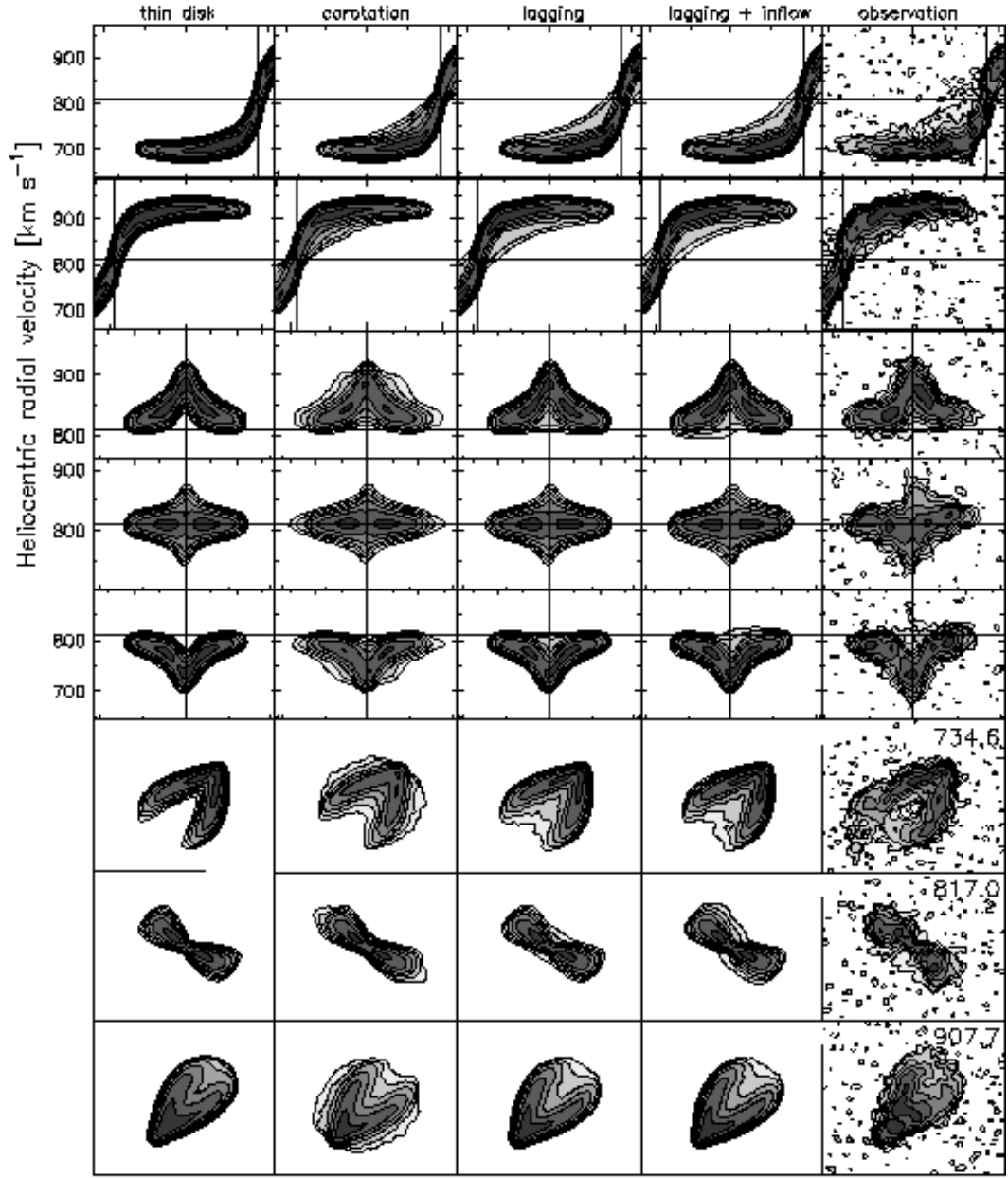}
\caption{Comparison of the H\,{\small I} observations of
NGC\,4559 with different models. The rightmost column shows the
data at 26$\arcsec$ resolution. From the top: position-velocity along the major axis on the
approaching side and on the receding side, parallel to the minor
axis (+1$\arcmin$ N-W, minor axis, -1$\arcmin$ S-E) and three channel maps (the
heliocentric radial velocities are on the top
right  of the data frames). Contours are -2, 2, 4, 8, 16, 32, 64
$\sigma$, with $\sigma$ = 0.48 mJy beam$^{-1}$. From the left: one-component
model made of a thin (scale height = 0.2 kpc) disk (thin
disk); two-component model made of a thin
disk and a co-rotating thicker (scale height 4 kpc) layer (corotation); two-component
model made of a thin disk and a slowly rotating thicker (scale height 2 kpc) layer (lagging)
and with a radial motion toward the centre of the galaxy (lagging + inflow).}
\label{models}
\end{figure}

First of all we have to understand if the
anomalous gas has a truly anomalous kinematics or if the anomalies
that are found are produced by projection effects due to the
inclination and/or thickness of the H\,{\small I} disk. We have
done this with the aid of 3D models made with the task GALMOD
(GIPSY) modified by us to include radial motions and a
vertical density profile for the thick disk. We have considered
several variations of a two-component structure, made of a thin
and a thicker, but lower density layer. These two components do
not actually coexist in the same spatial region of the equatorial
plane. The input values for the rotation curve and the radial
H\,{\small I} density profile are those derived from the data. The
models have been inspected in the same way as the data cube to
make a full comparison with the data. Figure \ref{models} shows
the position-velocity diagrams along the major axis (upper two
rows), three cuts along and parallel to the minor axis (central
three rows), and three channel maps (lower three rows). The
observations are in the rightmost column. The leftmost column
shows the thin disk model (scale height 0.2 kpc). The scale
height of the thin disk used here is similar to that of the
neutral hydrogen layer of the Milky Way or of other galaxies seen
edge-on. Note, however, that the size of the regions that can be
resolved in the data is a factor $\geq$ 5 times larger than the
adopted scale height for the thin disk. Clearly, the thin disk
model fails to reproduce the tail at low rotation velocities in
the observed p-v diagram along the major axis.

We have tried to reproduce this tail by varying only the thickness
of the gas layer. To this end, we have constructed a two-component
structure (second column) made of a  thin disk (scale height 0.2
kpc) and a corotating thicker layer (scale height 4 kpc). This
model does reproduce part of the low level emission visible in the
position-velocity diagram along the major axis, but the p-v
diagrams parallel to the minor axis and the channel maps disagree
with the data. This shows that the anomalous gas in NGC\,4559
cannot be caused by thickness effects alone. The third
column shows a model in which the gas layer is composed of two
non-corotating components: a thin disk (scale height 0.2 kpc)
rotating with the rotation curve found for the cold disk (Fig.
\ref{rotcur_anom}) and a thick disk (scale height 2 kpc) rotating
with the rotation curve of the anomalous gas shown in Fig.
\ref{rotcur_anom}. This model provides a good representation of
the data showing that the anomalous gas is indeed produced by
a combination of lagging and thickness.

The velocity field of the anomalous gas (Fig. \ref{anom}) shows
that the projected kinematical minor axis (thick line) differs,
mainly in the inner part, from that of the cold disk (cf. Fig.~1).
In NGC\,2403 the projected kinematical minor axis appears to be
rotated with respect to that of the cold disk. This has been
explained in terms of an overall inflow motion of the extra-planar
gas towards the centre of the galaxy (Fraternali et al. 2002). We
have explored the possibility of an inflow motion of the anomalous
gas toward the centre also for NGC\,4559, by making a model
(column next to the data) equal to the previous one with the
addition of a radial motion toward the galaxy centre (inflow) of
15 km s$^{-1}$. There are some hints of radial motion also
in NGC\,4559, but the evidence is not as strong as for NGC\,2403.

\section{Discussion and conclusions}

\subsection{Holes in the H\,{\small I} distribution}
\label{holes}

\begin{figure}
\includegraphics[width=140mm]{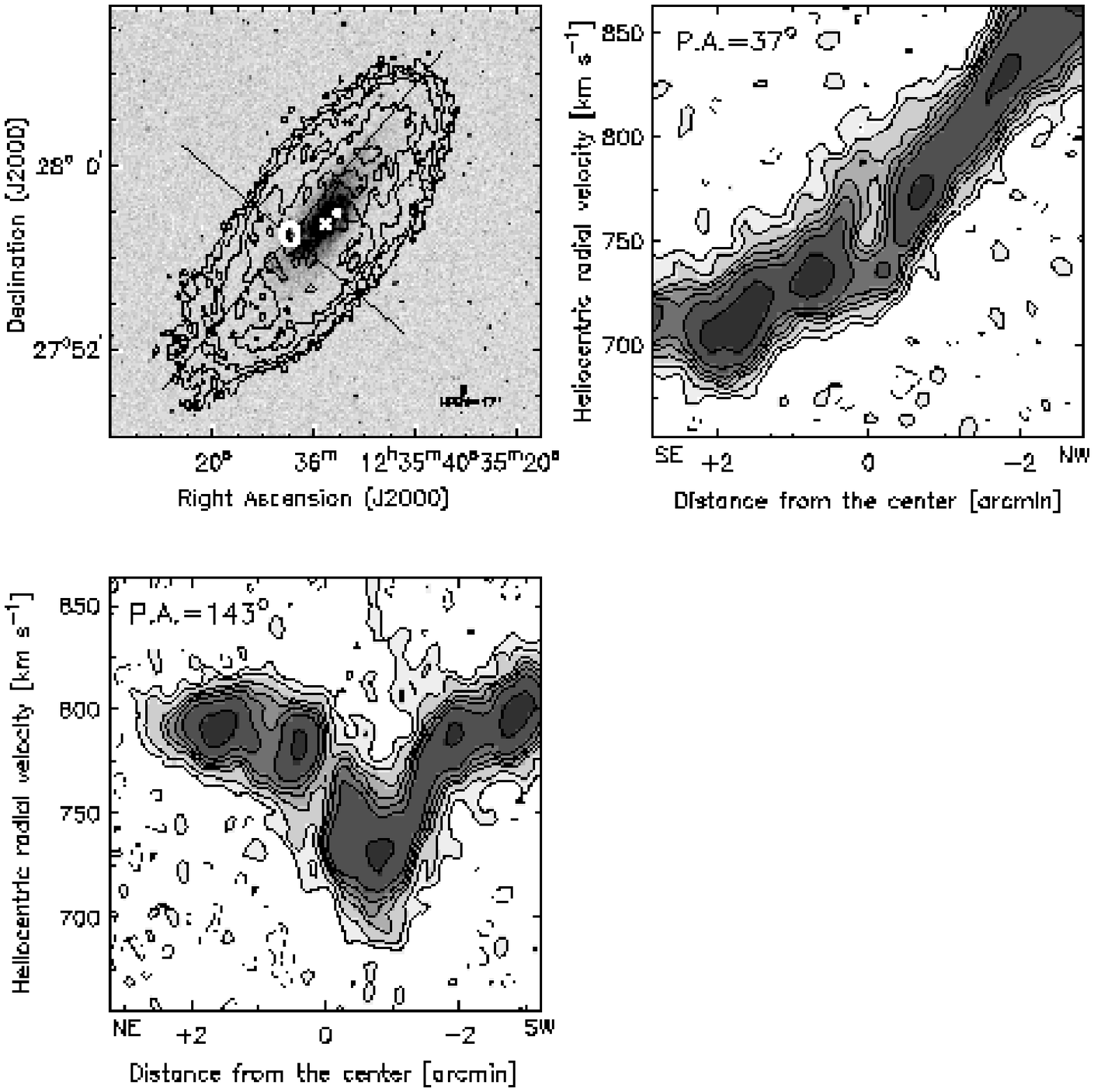}
\caption{Top left: Total H\,{\small I} column density distribution
  superposed on the optical image (DSS). The H\,{\small I} contours
  range from 1.5 $\times$ $10^{20}$ to 2.4 $\times$
  $10^{21}$ atoms cm$^{-2}$. The cross marks the position of the
  centre of the galaxy and the white dot indicates the position of SN
  1941. The cuts are parallel to the major and the minor axis and
  centered on the H\,{\small I} hole (white ellipse). Shown are also
  the corresponding position - velocity map (17$\arcsec$ resolution) parallel to the major axis (top right) and to the minor axis (bottom left). The contour levels are -1.5, 1.5, 3, 6, 8, 11, 16, 32 $\sigma$, with $\sigma$ = 0.52 mJy beam$^{-1}$. Negative contours are dashed.}
\label{buco15_art}
\end{figure}

The H\,{\small I} distribution shows several holes.
One of the most remarkable is located at $\alpha$ = 12$^h$36$^m$4$^s$
$\delta$ = 27$\degr$57$\arcmin$7$\arcsec$, a location marked by a
small white ellipse in Fig.\ref{buco15_art}.
The hole has a diameter of about 1 kpc.
It is especially clear in the position-velocity diagrams parallel to
the minor and to the major axis, centered on the hole, shown in
Fig. \ref{buco15_art}.
In the direction of the hole, we see a broadening and splitting of the
line profile.
This is the signature of an expanding shell of gas moving away from
the disk into the halo.
The systemic radial velocity of the shell is 760 km s$^{-1}$. If we
assume an isotropic expansion, the expansion velocity of the spherical
shell is about 30 km s$^{-1}$ as measured along the line of
sight.
However, the expansion of the shell might be anisotropic and faster in
the direction perpendicular to the disk, toward the low-density
halo.
In this case, we would be observing the projected component of the
expansion velocity which could then be as large as 80 km s$^{-1}$.

The hole is not completely empty, but contains a mass of about 3
$\times$ $10^6$ $M_{\sun}$. From the surrounding regions, we
have estimated an average column density of 2.5 $\times$ $10^{21}$
cm$^{-2}$, that would be expected if no hole were present. The mass
required to fill the hole is then $\sim$2 $\times$ $10^7$
$M_{\sun}$. This mass is comparable to the sum of the mass of gas
moving away from the disk ($\simeq$ 5 $\times$ $10^6$ $M_{\sun}$) and
the mass of the part of the shell presumably moving tangentially
($\simeq$ 7 $\times$ $10^6$ $M_{\sun}$). The hole and the shell may
have been produced by supernova explosions or stellar winds. An
estimate of the kinetic energy of the swept up gas gives values
between 2 $\times$ $10^{53}$ erg and 1 $\times$ $10^{54}$ erg.

If the expansion of the H\,{\small I} shell is isotropic, the
kinematic age of the hole lies between 6 $\times$ $10^6$ yr and
1.5 $\times$ $10^7$ yr. If the hole was initially spherical, the
differential rotation of the galaxy may be the cause of the observed
elongation of the hole in the direction of rotation. The time
necessary to produce such a stretching is of the order of $10^7$
yr. Another way to estimate the age of the hole is to use the velocity
dispersion ($\sim$ 10 km s$^{-1}$) of the gas surrounding the hole. One may
expect the hole to be filled in 5 $\times$ $10^7$ yr.

If holes and bubbles are produced by supernova explosions or stellar winds,
there should be some correlation with star formation in the
disk. Moreover the bubbles might be filled with hot gas (a few 10$^6$
K) which should be visible in the X-rays (Norman and Ikeuchi
1989). However, from an analysis of ROSAT observations of NGC\,4559,
Vogler, Pietsch and Bertoldi (1997) did not find any excess X-ray emission in the region of the
hole. Therefore, there is no evidence for the existence of hot
gas in the H\,{\small I} hole. If the kinetic energy is of the order of a few $10^{53}$ erg, a hundred to a thousand supernovae would be required to
produce the hole. We note that the hole is located within the optical
disk where the star formation activity is high. A Type II supernova
explosion was observed in NGC\,4559 in 1941 (SN 1941, Kowal and
Sargent 1971), at $\alpha$ = 12$^h$35$^m$55.5$^s$ $\delta$ =
27$\degr$58$\arcmin$0$\arcsec$ (see the small white dot in Fig. \ref{buco15_art}), far away from the
hole.

\subsection{Extra-Planar gas and the lopsidedness of NGC\,4559}
\label{origin}

High sensitivity observations of NGC\,4559 have revealed the
presence of  ``anomalous'' gas rotating more slowly than the gas
in the disk. 3D models have shown that this anomalous gas is
indeed a slowly rotating extra-planar component. In Sections
2 and 3 we have shown that the gaseous disk of NGC\,4559 is
lopsided both in density distribution and in kinematics. Here we
investigate the possible lopsidedness of the extraplanar gas by
constructing separate models for the approaching and receding
sides of the galaxy. 

\begin{figure}
\includegraphics[width=140mm]{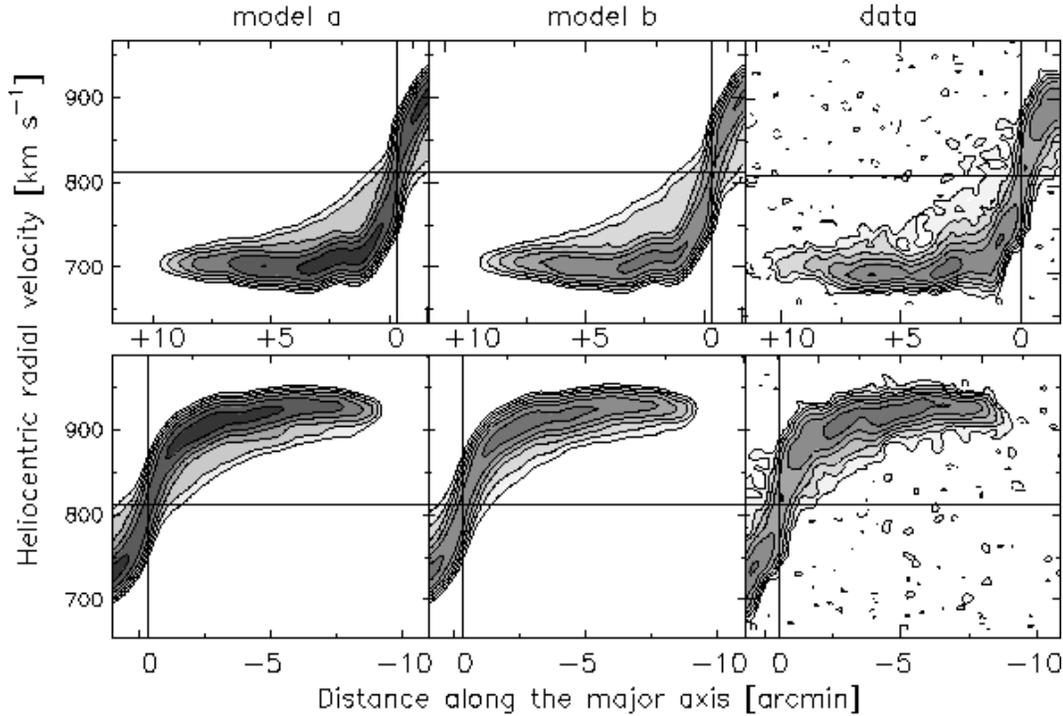}
\caption{Comparison of the H\,{\small I} observations of
NGC\,4559 with different models.
Position-velocity map along the major axis for the approaching side
(top) and for the receding side (bottom).
The rightmost column shows the data at 26$\arcsec$ resolution.
Contours are -2, 2, 4, 8, 16, 32, 64 $\sigma$, with $\sigma$ = 0.48
mJy beam$^{-1}$.
First column: two-component model made of a thin disk (scale height
0.2 kpc) and a slowly rotating thicker (scale height 2 kpc) layer
(model a).
Second column: two-component model made of a thin disk (scale height
0.2 kpc) and a slowly rotating layer with a scale height of 3 kpc on
the approaching side and of 1 kpc on the receding side (model b).}
\label{2comp_best}
\end{figure}

Fig. \ref{2comp_best} shows the observed position-velocity diagram
along the major axis on the approaching and the receding sides
(rightmost column).
The extra-planar gas component appears to be more pronounced on the
approaching side of the galaxy.
This may simply mean that the less steep rotation of the receding part
of the galaxy conspires to partly hide the anomalous gas on that
side.
The first column of Fig. \ref{2comp_best} shows the model
position-velocity diagrams for the two parts of the galaxy, with a
thin disk (scale height 0.2 kpc) and a more slowly rotating thicker
layer (scale height 2 kpc).
The rotation curves and the radial H\,{\small I} density profiles used
for the thin disk are those derived separately for the approaching and
for the receding sides, whereas the thicker layer is kept
symmetrical.
The observed asymmetry is reproduced, but not completely.
On the approaching side, the model reproduces less anomalous gas than
observed, while on the receding side it seems to produce slightly
more.
A somewhat better reproduction of the asymmetry can be obtained in two
different ways.
The first is to consider different rotation curves of the extra-planar
gas on the two sides of the galaxy, lower on the approaching and
higher on the receding side. The second is to make the thick disk
thicker on the approaching and thinner on the receding side.
In model b of Fig.\ref{2comp_best} we have done this.
We show here a model composed of a thin disk (scale height 0.2 kpc)
and a more slowly rotating thicker layer.
For the thin disk we have used the same input values of the previous
model whereas we have made the thick disk more massive and thicker on
the approaching side (scale height 3 kpc against 1 kpc).
This model provides a good representation of the data.
Here, the rotation curve of the thick layer has been kept
symmetrical.
If, however, the H\,{\small I} layer on the approaching side is indeed
thicker, its rotation velocity may be lower as found for NGC\,891 by
Fraternali et al. (2004).

In conclusion, it seems that the density distribution and the
kinematics of the extraplanar gas in NGC\,4559 are not azimuthally
symmetric. The extraplanar gas seems more abundant on the
approaching side. In the optical and radio continuum images
(Fig. \ref{pres}), this side appears brighter than the receding one and
possibly has, therefore, higher star formation activity (see also
the X-ray image in \cite{cro04}). Moreover, this is the side where
the rotation curve rises more steeply indicating a higher
concentration of mass. The fact that the extraplanar gas seems
prominent on this side supports the idea of a link between
extraplanar gas and star formation.

Another indication of the connection with star formation comes
from the presence of the forbidden gas, i.e.\ the gas  seen in
Fig. \ref{mod_cod} and Fig. \ref{2comp_best} near the centre
(0$\arcmin$ to $+$2$\arcmin$.5, $2^{nd}$ quadrant) at
``forbidden'' velocities differing by about 150 km s$^{-1}$ from
the rotation velocity. This gas is not reproduced by our model
probably because we have not considered vertical motions. It is
interesting to note that the forbidden component is found only in
the central regions (within 5 kpc) of the galaxy and mainly on the
approaching side. Moreover, this region is close to the remarkable
hole in the H\,{\small I} distribution described in Sect.
\ref{holes}. 

\subsection{The origin of the extra-planar gas}

The origin of the extra-planar gas is not known.
The main question is whether it is caused by processes that take
place in the galactic disk or whether it originates from infall of
extragalactic, primordial gas.

With this study we have collected new evidence for a close
relation between the extra-planar gas and the star formation
activity: 1) the overall kinematical pattern of the extraplanar
gas is regular and suggests a close connection with the underlying
thin disk of NGC\,4559 and a galactic fountain origin; 2) the
extraplanar gas seems more abundant on the brighter (approaching)
side of the galaxy; 3) there is evidence for fast moving gas
complexes (forbidden gas) and large expanding HI supershells.
Moreover, 4) the extra-planar gas appears to be regularly
distributed over the whole galaxy. This suggests that the
extra-planar gas is unlikely to be due to a local phenomenon, such
as the recent accretion of a massive cloud, and it is more likely
to be the result of a widely spread activity such as star
formation. Indeed, observations in other bands provide evidence
for a regularly distributed star formation activity in the disk of
NGC\,4559 (e.g.\ \cite{cro04}).

However, despite all these arguments apparently favoring an
internal origin of the extra-planar gas, an interpretation of the
observed phenomena in terms of accretion from the intergalactic
medium cannot be ruled out completely. In fact, the above
arguments, which do suggest a link between the extra-planar gas
and star formation in the disk, could be reversed to conclude that
the higher star formation rate might be the end result of
accretion rather than the cause at the origin of extra-planar gas.

It is also worth mentioning here that, using X-ray and HST optical
data, \cite{sor05} claim to have found evidence for a recent
collision of a satellite dwarf with the disk of NGC\,4559.

In conclusion, the strongest point in favor of an internal origin
seems to be the kinematical pattern of the extra-planar gas. It is
very difficult to produce such a regular pattern from accretion of
external material having, in principle, randomly oriented angular
momentum. This suggests that at least a significant fraction of
the extra-planar gas is likely to originate from galactic
fountains.

The extra-planar gas of NGC\,4559 is similar to that found by
Swaters et al. (1997) in the edge-on spiral galaxy NGC\,891, by
Matthews and Wood (2003) in UGC\, 7321, and by Fraternali et al.
(2002) in NGC\,2403. The parameters for the thick disk derived in
these studies are similar. For instance, the extraplanar gas in
NGC\,2403 has a mass of 1/10 of the total H\,{\small I} mass, a
mean rotation velocity of 25-50 km s$^{-1}$ lower than that of the
disk, and a radial inflow of about 10-20 km s$^{-1}$ towards the
galaxy centre. The H\,{\small I} observations presented here are,
together with those of NGC\,2403, NGC\,891 and UGC\,7321, among
the deepest ever obtained. This suggests that the extra-planar gas
may be a common feature in spiral galaxies, missed earlier because
of insufficient sensitivity of the previous observations. Clearly,
new observations are required to confirm this interesting
possibility and to study the relations of the extra-planar gas
with the galaxy morphological type, star formation rate, and
environment. 

\begin{acknowledgements}
Claudia Barbieri is grateful to ASTRON (Dwingeloo), to the Kapteyn
Astronomical Institute (Groningen), to the INAF-Osservatorio
Astronomico di Bologna and to the Istituto di Radioastronomia
(CNR, Bologna) for their hospitality and financial support. The
Westerbork Synthesis Radio Telescope is operated by the ASTRON
(Netherlands Foundation for Research in Astronomy) with support
from the Netherlands Foundation for Scientific Research NWO.
\end{acknowledgements}

\appendix

\section{Mass model}
\label{mass model}

\begin{figure}
\includegraphics[width=140mm]{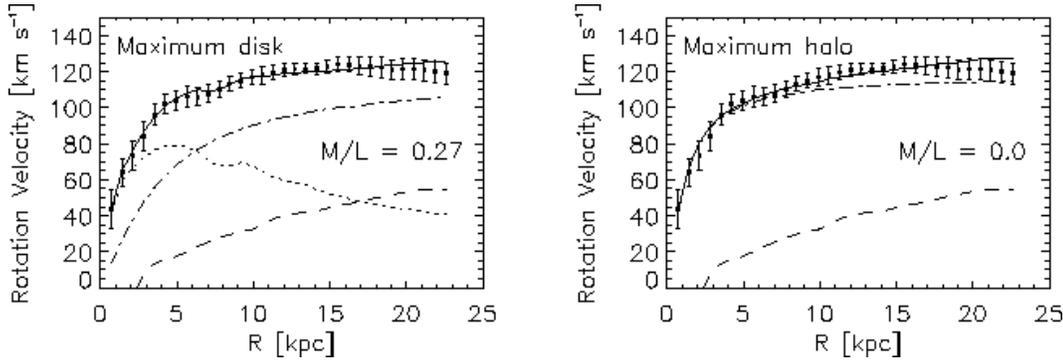}
\caption{Two different mass models for NGC\,4559. The filled
squares show the observed rotation curve. The contributions of the
gas (dashed), stars (short dashed) and dark matter halo
(long-short dashed) are shown. The values of the mass-to-light
ratio are for the $K$ band.} \label{isot}
\end{figure}

The rotation curve shown in Fig. \ref{rotcur} serves as the basis for the study of the galaxy dynamics. The adopted mass model consists of  three components (Begeman 1989): the stellar disk, the gas disk, and
the dark matter halo. For the
stellar component we have used the surface brightness
profile in $K$-band obtained from the 2MASS catalog (Jarrett et al. 2003), fitted with an exponential law for radii larger than 0.8 kpc. The
derived scale length is 1.9 kpc and the central brightness is
17.14 mag/arcsec$^2$. Figure \ref{global_rev} (right panel) shows the radial column
density profile for the neutral hydrogen. This has been obtained from the observed total H\,{\small
I} map, using the kinematical parameters from the tilted
ring model. The gas column density is obtained by multiplying the H\,{\small I} density by a factor
1.4 to take into account the helium content. Note
that NGC\,4559 shows a central depression in the hydrogen
distribution ($R$ $\leq$ 2 kpc). Since the force field is calculated
from the observed H\,{\small I} surface density, the potential of
the gaseous disk may be such that for the innermost regions there
is a net outward pull, resulting formally in imaginary rotation velocities.
For this reason, the velocity contribution of the gaseous disk in
Fig. \ref{isot} starts from a radius of about 2 kpc. For the scale
heights, we have assumed a value of 0.4 kpc for the stellar
and 0.2 kpc for the gas layer.
The velocity contribution of the luminous disk
(star + gas) has been obtained from the formula given by
Casertano (1983). For the dark matter halo we have adopted a
spherically symmetric density profile $\rho(R) = \rho_0 (1 +
R/R_0)^{-2}$, where $\rho_0$ is the central density and $R_0$ is
the core radius.

Figure \ref{isot} shows two different fits to the rotation curve.
The maximum-disk solution is presented in the left panel. The
corresponding value of the mass-to-light ratio, obtained using the photometric profile in $K$ band, is 0.27
$M_{\sun}/L_{K,\sun}$. This value includes the contribution of the
molecular gas component, based on the assumption that this follows the light distribution (Young and Knezek 1989).
The second model is the so-called maximum halo, in which formally
no stellar disk is considered (0.0 $M_{\sun}/L_{K,\sun}$). The
quality of the fits are comparable. Analyzing the sample of Ursa Major galaxies, Verheijen (1997)
finds that the $K'$ mass-to-light ratios of the High Surface
Brightness galaxies under the maximum disk hypothesis typically range from 0.4 to 1.0, slightly higher than the value found here.

\end{document}